\documentclass[letterpaper,twocolumn,notitlepage]{revtex4-1}
\usepackage{amsmath,amssymb}
\usepackage[pdftex]{graphicx}
\usepackage{newtxtext,newtxmath}
\usepackage{bm,wasysym,mathtools}
\usepackage{color}
\usepackage[pdftex,unicode,colorlinks,bookmarks=true]{hyperref}

\definecolor{darkred}{rgb}{0.7,0,0}

\renewcommand{\Im}{\mathop{\rm Im}\nolimits}

\begin{document}

\title{Overcoming the SQL in gravitational wave detectors using spin systems with negative effective mass}
 
\author{F.Ya.Khalili}

\email{khalili@phys.msu.ru}

\affiliation{Faculty of Physics, M.V. Lomonosov Moscow State University, 119991 Moscow, Russia}

\affiliation{Russian Quantum Center, Skolkovo 143025, Russia}

\author{E.S.Polzik}

\email{polzik@nbi.ku.dk}

\affiliation{Niels Bohr Institute, University of Copenhagen, 2100, Denmark}



\begin{abstract}
Quantum back action (QBA) of a measurement limits the precision of observation of the motion of a free mass. This profound effect dabbed the "Heisenberg microscope" in the early days of quantum mechanics, leads to the standard quantum limit (SQL) \cite{92BookBrKh} stemming from the balance between the measurement sensitivity and the QBA. Here we consider the measurement of motion of a free mass performed in a quantum reference frame with an effective negative mass which is not limited by QBA. As a result, the disturbance on the motion of a free mass can be measured beyond SQL. QBA-limited detection of motion for a free mass is extremely challenging, but there are devices where this effect is expected to play an essential role, namely, gravitational wave detectors (GWD) such as LIGO and VIRGO. Recent reports on observation of gravitational waves \cite{PRL_116_061102_2016, PRL_116_241103_2016, PRL_118_221101_2017, PRL_119_141101_2017, PRL_119_161101_2017} have opened new horizons in cosmology and astrophysics. Here we present a general idea and a detailed numerical analysis for QBA-evading measurement of the gravitational wave effect on the GWD mirrors which can be considered free masses under relevant conditions. The measurement is performed by two entangled beams of light probing the GWD and an auxiliary atomic spin ensemble, respectively. The latter plays a role of a free negative mass. We show that under realistic conditions the sensitivity of the GWD can be significantly increased over the entire frequency band of interest.
\end{abstract}

\maketitle

\paragraph{Introduction}

Position of a mass as a function of time, $\hat{x}(t)=\hat{x}(0)+\hat{\dot{x}}(0)t=\hat{x}(0)+\hat{p}(0)t/m$, cannot be determined precisely because $\hat{x}(0)$ and $\hat{p}(0)$ are non-commuting operators and if one of them is measured, the other one is disturbed by QBA. However, if the position is measured relatively to a reference frame associated with another quantum system described by $\hat{x}_0(0)$ and $\hat{p}_0(0)$ and negative effective mass $-m$, then $\hat{x}(t)-\hat{x}_0(t)=\hat{x}(0)-\hat{x}_0(0)+(\hat{\dot{x}}(0)-\hat{\dot{x}}_0(0))t=\hat{x}(0)-\hat{x}_0(0)+(\hat{p}(0) + \hat{p}_0(0))t/m$. In this case, the relative position is defined by commuting operators $\hat{x}(0)-\hat{x}_0(0)$ and $\hat{p}(0)+\hat{p}_0(0)$ and hence can be QBA-free. Narrowband QBA evading methods for high frequency  oscillators have been proposed and utilized experimentally \cite{Julsgaard_Nature_413_400_2001, Hammerer_PRL_102_020501_2009, Zhang_PRA_88_043632_2013, Wooley_PRA_87_063846_2013, Ockeloen-Korppi_PRL_117_140401_2016, Moeller_Nature_547_191_2017}. In \cite{Tsang_PRL_105_123601_2010, Tsang_PRX_2_031016_2012} the negative mass approach has been analyzed from an information-theoretical perspective.

In this Letter we propose an idea and present a detailed analysis for application of the measurement of a free mass in a negative mass reference frame to gravitational wave detectors (GWDs). Due to the unique technical parameters of the GWDs, their massive mirrors can be considered free masses in the relevant range of time and frequency. At the same time sensitivity of state-of-the-art laser interferometric GWDs, such as Advanced LIGO \cite{LIGOsite}, Advanced VIRGO \cite{VIRGOsite}, and GEO600 \cite{GEOsite} is to a major extent limited by quantum fluctuations of the probing light. 

For short measurement times (a higher frequency range) the sensitivity of measurement of motion is dominated by the phase {\it shot noise} of light.  This noise can be suppressed \cite{Caves1981} by using a phase-squeezed state of light, as demonstrated at the GEO600 GW detector \cite{Nature_2011, Grote_PRL_110_181101_2013}. However, suppression of the phase fluctuations leads to proportional increase of the amplitude fluctuations, or the  {\it radiation pressure noise} which is the origin of the QBA noise.  QBA dominates for longer measurement times, as 
its intensity is proportional to the free mass mechanical susceptibility of suspended mirrors of GW detectors, $\chi=-1/\Omega^2$. 
For the Advanced LIGO, the radiation pressure (QBA) noise will be a major limitation \cite{CQG_32_7_074001_2015}.  

Suppression of both the shot noise and the QBA, and thus overcoming the SQL, requires more advanced methods than ordinary frequency-independent squeezing. Methods proposed towards this goal to-date \cite{Chen2002, 02a1KiLeMaThVy, Ma_NPhys_13_776_2017, 12a1DaKh} are challenging as they involve large scale installations and/or modifications of the GWD core optics. 


Here we show that a suitably designed atomic spin ensemble provides a reference frame in which a broadband quantum noise reduction for motion of free masses, such as the GWD mirrors, is possible. 

\begin{figure}
  \includegraphics[width=\columnwidth]{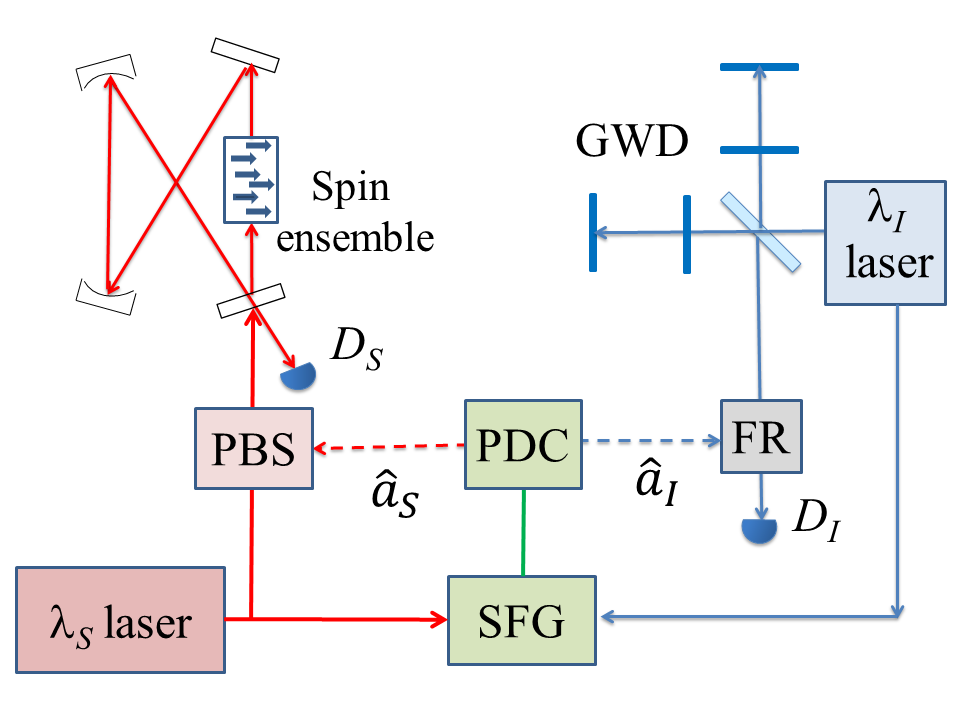}
  \caption{\textbf{Setup for a GWD beyond the SQL with the negative mass spin system}. The GWD and the atomic system are probed with entangled light modes $a_S, a_I$ (dashed lines). The modes are generated through sum frequency generation (SFG) of the GWI laser and an auxiliary laser at the atomic frequency $\lambda_S$, and the subsequent parametric downconversion (PDC). Combined signals from detectors $D_S$ and $D_I$ allow for back action free measurement. PBS --- polarization beam splitter; FR --- Faraday rotator.}\label{Fig:setup}
\end{figure}

\paragraph{The scheme.}

The schematic of the proposed experimental realization for detection of the free mass motion in the negative mass reference frame is presented in Figure \ref{Fig:setup}. Two quantum measurements are performed in parallel, the measurement of the position of the end mirrors of the GW interferometer with the optical field $\hat{{\rm a}}_I$, and the measurement on the auxiliary atomic spin ensemble with the field $\hat{{\rm a}}_S$. The two fields are centered at wavelengths $\lambda_I$ and $\lambda_S$, respectively, where $\lambda_I$ is determined by the probing laser of the GWD (presently $1064$nm) and $\lambda_S$ - by an atomic resonant transition. As we shall show below if  $\hat{{\rm a}}_I$ and  $\hat{{\rm a}}_S$ are in an entangled state, both the shot noise and the radiation pressure (QBA) noise contributions to the joint measurement on the two systems can be suppressed.


In the absence of optical losses, with the interferometer tuned on resonance, the phase quadrature of the light mode exiting the interferometer, $\hat{{\rm b}}_1^s$, measured by a homodyne detector $D_I$ is \cite{Buonanno2003, 12a1DaKh}:
\begin{equation}\label{b_I_s} 
  \hat{{\rm b}}_I^s = \frac{\kappa_I+i\Omega}{\kappa_I-i\Omega}\,\hat{{\rm a}}_I^s 
    + \frac{2\kappa_I\Theta\chi}{(\kappa_I- i\Omega)^2}\,\hat{{\rm a}}_I^c 
    + \frac{\sqrt{2\kappa_I\Theta}}{\kappa_I- i\Omega}\,\chi\frac{F_s+F_T}{\sqrt{\hbar m}}
  \,,
\end{equation} 
where $\hat{{\rm a}}_I^s$, $\hat{{\rm a}}_I^c$ are the phase (sine) and amplitude (cosine) quadratures of the incident light, $F_s$ is the signal force, for example from GW, and $F_T$ is a sum of the thermal force, seismic noise and other technical noise sources (notations used throughout this paper are listed in Table \ref{table:notations}). The first term describes the shot noise and the second one --- the QBA noise. 

\begin{table*}
  \begin{ruledtabular}
    \begin{tabular}{llll}
      Notation & Quantity & Value, Adv.LIGO & Value, 10-m \\
      \hline
      $r$ & Squeezing factor & \multicolumn{2}{c}{$\dfrac{\log 15}{2} \leftrightarrow 15\,{\rm db}$} \\
      $L$ & Interferometer arms length  & 4000\,m & 10\,m \\
      $m$ & Mirrors mass & 40\,kg & 0.1\,kg \\
      $\kappa_I$ & Interferometer half-bandwidth & $2\pi\times500\,{\rm Hz}$ & $2\pi\times2000\,{\rm Hz}$ \\
      $I_c$ & Optical power circulating in each of the arms & 840\,kW & 1\,kW \\ 
      $\Theta = \dfrac{8\omega_oI_c}{mcL}$ & normalized optical power & $(2\pi\times100)^3\,{\rm s}^{-3}$ & $(2\pi\times575)^3\,{\rm s}^{-3}$\\
      $\Omega_S$ & Atomic system eigen frequency & $2\pi\times3\,{\rm Hz}$ & $2\pi\times30\,{\rm Hz}$ \\
      $\gamma_S$ & Atomic system damping rate & $2\pi\times3\,{\rm Hz}$ & $2\pi\times30\,{\rm Hz}$ \\
    \end{tabular}
  \end{ruledtabular}
  \caption{The main notations used throughout this paper}\label{table:notations} 
\end{table*}

If the incident light is in a coherent or in a squeezed state, then the quadratures $\hat{{\rm a}}_I^s$, $\hat{{\rm a}}_I^c$ are uncorrelated and their spectral densities are equal to $e^{-2r}/2$ and $e^{2r}/2$. It is easy to show \cite{Caves1981, 12a1DaKh} that in this case the spectral density of the sum of the shot noise and QBA quantum noise (normalized to to signal force $F_s$) cannot be smaller than the force SQL $S^F_{\rm SQL} = \hbar m\Omega^2$. Typically, it is recast as the equivalent position SQL:   
\begin{equation}
  S^x_{\rm SQL} = \frac{S^F_{\rm SQL}}{(m\Omega^2)^2} = \frac{\hbar}{m\Omega^2} \,.  
\end{equation} 

\begin{figure}
  \includegraphics[width=\columnwidth]{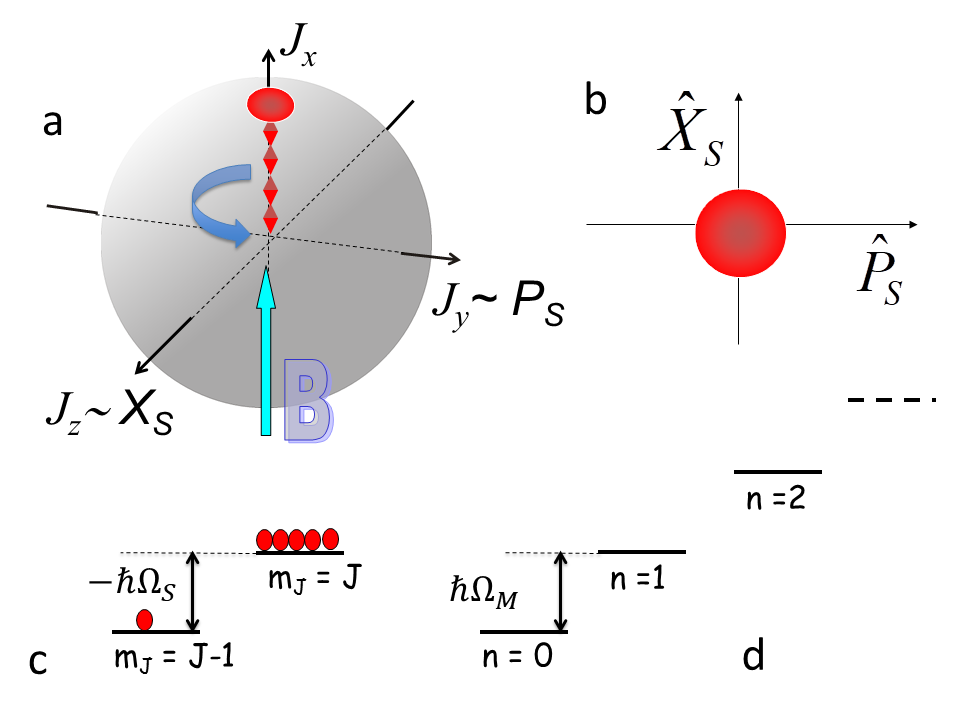}
  \caption{\textbf{Macroscopic spin oscillator}. a). A collective spin of an ensemble of atoms in magnetic field shown as a vector on a Bloch sphere. b) Normalized spin components orthogonal to the mean spin direction are equivalent to canonical operators. c). For the geometry shown in a) the first excited state of the spin oscillator has the energy which is below the state with no excitations corresponding to the negative mass oscillator. d). Mechanical oscillator spectrum.}\label{Fig:spin}
\end{figure}

Let us now introduce the second quantum system consisting of a multi-atom spin ensemble. If the spins are optically polarized along a certain direction $x$ (Figure \ref{Fig:spin}) the collective spin has a large average projection $J_x=|\langle \hat{J}_x \rangle|/\hbar \gg 1$ \cite{Hammerer_PRL_102_020501_2009, Hammerer_RMP_82_1041_2010}. Its normalized $y,z$ quantum components form canonical variables $\hat{X}_S = \hat{J}_z/\sqrt{\hbar J_x}$, $\hat{P}_S=-\hat{J}_y/\sqrt{\hbar J_x}$, satisfying the commutation relation $[\hat{X}_S,\hat{P}_S]=i$. In terms of those variables, the Hamiltonian for the ensemble placed in magnetic field oriented along $x$ becomes
\begin{equation}
  \hat{H}_{S}= \hbar\Omega_SJ_x - \frac{\hbar\Omega_S}{2}(\hat{X}_{S}^2+\hat{P}_{S}^2) \,,
\end{equation} 
where $\Omega_S$ is the Larmor frequency. The first term is an irrelevant constant energy offset due to the mean spin polarization. The second term is equivalent to the Hamiltonian of a mechanical oscillator $\hat{H}_M$ with a \textit{negative} mass and spring constant. Each quantum of excitation in the negative mass spin oscillator physically corresponds to a deexcitation of the inverted spin population from its highest energy level by $\hbar\Omega_S$. Preparation of the collective spin in the energetically lowest Zeeman state realizes instead a \textit{positive} mass and spring constant spin oscillator.

Interaction of light with a spin in magnetic field placed inside a resonator with the finess $\mathcal{F}$ can be cast in the form similar to that for the mechanical oscillator (see Supplementary Information):
\begin{equation}\label{spin_noise}
  \hat{{\rm b}}_S^s = \hat{{\rm a}}_S^s + 2\theta\chi_S\hat{{\rm a}}_S^c
    + \sqrt{2\theta}\,\chi_S\hat{f}_S \,,
\end{equation}
where $\hat{{\rm a}}_S^s$, $\hat{{\rm a}}_S^c$ are the phase (sine) and amplitude (cosine) quadratures of the input light mode in polarization orthogonal to the linearly polarized driving optical field, $\theta = \Omega_S\Gamma_S$, $\Gamma_S = \gamma_Sd_0$ is the spin oscillator read out rate, $d_0=\frac{2\mathcal{F}}{\pi}\frac{\sigma N_a}{\mathcal{A}}$ is the cavity enhanced resonant optical depth of the spin ensemble, $\gamma_S=\frac{\sigma}{A}\frac{\gamma_{\rm opt}^2\Phi}{\Delta_{\rm opt}^2}$ is the spin bandwidth dominated by the optically induced decoherence, $N_a$ is the atoms number, $\sigma$ - the atomic optical crossection, $\mathcal{A}$ --- the spin ensemble crossection, $\gamma_{\rm opt}$ --- optical transition bandwidth, $\Delta_{\rm opt}$ - optical field detuning from atomic resonance, $\Phi$ --- photon flux, 
\begin{equation}\label{chi_S} 
  \chi_S = -[(\gamma_S-i\Omega)^2 + \Omega_S^2]^{-1} 
\end{equation}
is the effective susceptibility of the spin oscillator, and $\hat{f}_S$ is a normalized thermal force acting on the spin \cite{Hammerer_RMP_82_1041_2010, Moeller_Nature_547_191_2017}. An interesting and useful feature of a spin oscillator is that it is possible to provide the effective temperature of the noise $\hat{f}_S$ close to zero even if the collective spin is formed by a gas of atoms at room temperature \cite{Hammerer_RMP_82_1041_2010}. Under such conditions the spectral density of this noise force corresponds to zero point fluctuations:
\begin{equation}
  S_S = |\Im\chi_S^{-1}| = 2|\Omega|\gamma_S \,.
\end{equation} 


\begin{figure}
  \includegraphics[width=\columnwidth]{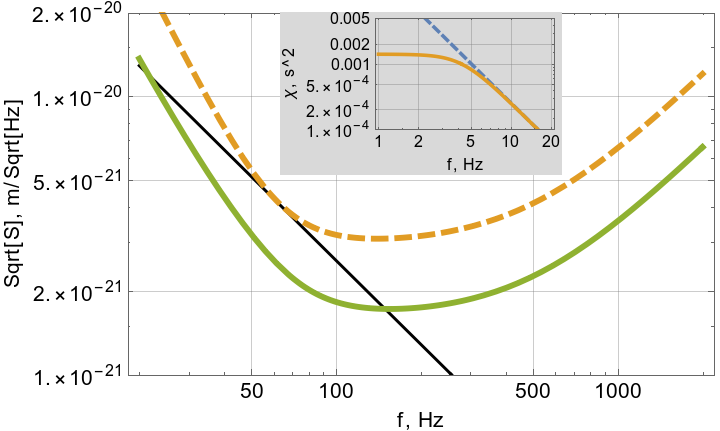}\\
  \includegraphics[width=\columnwidth]{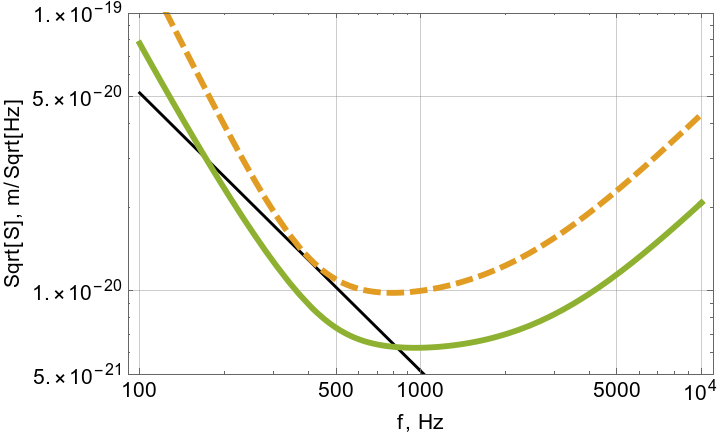}
  \caption{\textbf{Spectral densities of quantum noise}. Thin solid line - SQL; dashed orange curve - SQL-limited GWD noise; solid green curve - hybrid GWD/spin system. Top: Advaced LIGO, bottom: 10-m prototype (see Table \ref{table:notations}). In all cases, 2.5\% of input losses and 2.5\% of output losses are assumed for both GWD and atomic spin channels. In addition, 0.01\% of intracavity roundtrip losses for the GWD and 0.3\% for the spin system are assumed. Inset: dashed line --- susceptibility function $\chi$ for a free mass; solid line --- absolute value of susceptibility of the spin system with $\Omega_S=\gamma_S=2\pi\times3\,{\rm Hz}$.}
\label{Fig:noise}
\end{figure}


The sine quadratires $\hat{{\rm b}}_{I,S}^s$ of the output beams are measured by the homodyne detectors $D_{I,S}$, respectively. Let us first consider an ideal case within the frequency band $\Omega\ll\kappa_I$ when the detectors' outputs are added to measure
\begin{multline}
  \hat{{\rm b}}_I^s + \hat{{\rm b}}_S^s
  = \hat{{\rm a}}_I^s + \hat{{\rm a}}_S^s
    + \frac{2\Theta\chi}{\kappa_I}\hat{{\rm a}}_I^c+2\theta\chi_S\hat{{\rm a}}_S^c \\
    + \sqrt{\frac{2\Theta}{\kappa_I}}\,\chi\frac{F_s+F_T}{\sqrt{\hbar m}} 
    + \sqrt{2\theta}\,\chi_S\hat{f}_S\tanh2r \,.
\end{multline}
If the probe fields are perfectly entangled such that $\hat{{\rm a}}_I^s =- \hat{{\rm a}}_S^s$ and $\hat{{\rm a}}_I^c = \hat{{\rm a}}_S^c$, the response of the spin system is characterized by the effective negative mass and matches the response of the interferometer
\begin{equation}\label{QND_cond} 
  \frac{\Theta\chi}{\kappa_I} = -\theta\chi_S \,,
\end{equation} 
then \textit{both the phase shot noise and the QBA} are cancelled in the combined measurement record,  resulting in the measurement sensitivity not limited by SQL.


Entangled probe fields can be prepared by nonlinear optical transformations (sum frequency generation and parametric down conversion) as shown in Figure \ref{Fig:setup} \cite{Walls_Milburn2008, Schori_PhysRevA_66_033802_2002}. A small fraction of the GW detector laser and a laser locked to an atomic transition generate a pump beam through the sum frequency generation in $\chi^{(2)}$ medium. This beam is then used to pump a parametric downconversion process (PDC) in which two-mode squeezed vacuum modes $\hat{{\rm a}}_{I,S}$ at wavelengths $\lambda_1$ and $\lambda_2$ are generated  \cite{Schori_PhysRevA_66_033802_2002, Georgiades_PRL_75_3426_1995} satisfying
\begin{subequations}
  \begin{gather}
    \hat{{\rm a}}_{I,S}^c = \hat{{\rm z}}_{I,S}^c\cosh r + \hat{{\rm z}}_{S,I}^c\sinh r
      \,,\\ 
    \hat{{\rm a}}_{I,S}^s = \hat{{\rm z}}_{I,S}^s\cosh r - \hat{{\rm z}}_{S,I}^s\sinh r\,.
  \end{gather}
\end{subequations}
Here $\hat{{\rm z}}_1^{c,s}$ and $\hat{{\rm z}}_2^{c,s}$ correspond to two independent vacuum fields.

 For a finite degree of entanglement, the outputs of the detectors $D_S$ and $D_I$ should be added with the optimal weigth function, to obtain the readout with the suppressed quantum noise. It is straightforward to show using Eqs.\,(\ref{b_I_s}, \ref{spin_noise}),  that in the  lossless case,  within the bandwidth of the interferometer, $|\Omega|\ll\kappa_I$, (where the QBA is significant) the weigth factor is equal to $\tanh2r$. In this case the combined output is equal to
\begin{multline}
  \hat{{\rm b}}_I^s + \hat{{\rm b}}_S^s\tanh2r 
  = \frac{\hat{{\rm z}}_I^s\cosh r + \hat{{\rm z}}_s^s\sinh r}{\cosh2r} \\
    + \frac{2\Theta\chi}{\kappa_I}\,
        \frac{\hat{{\rm z}}_I^c\cosh r - \hat{{\rm z}}_s^c\sinh r}{\cosh2r} \\
    + \sqrt{\frac{2\Theta}{\kappa_I}}\,\chi\frac{F_s+F_T}{\sqrt{\hbar m}} 
    + \sqrt{2\theta}\,\chi_S\hat{f}_S\tanh2r \,,
\end{multline}
which corresponds to the noise supprression factor for both the shot noise and the QBA noise equal to $\cosh2r$. Importantly, the lasers $\lambda_I$ and $\lambda_S$ do not have to be phase locked to each other. Rather only the phases between the local oscillators and the respective laser beams should be stabilized, so that the correct quadratures are detected.

\paragraph{Numerical estimates.}

The quantum noise spectral density of the considered scheme is calculated in the SI, taking into account the optical losses in the interferometer and the spin system for two sets of parameters, one of which approximately corresponds to the design goals of the Advanced LIGO \cite{CQG_32_7_074001_2015} and the other one --- to the Hannover 10-m prototype interferometer \cite{10m_site, Westphal_1111_7252}, see Table \ref{table:notations}.   We show that the spin system allows for a broadband sensitivity beyond the SQL for both interferometers.

The critical parameters of the spin system are deduced from condition \eqref{QND_cond} using the parameters of the corresponding GWDs and of the atomic spin system reported in \cite{Moeller_Nature_547_191_2017}.  Consider, for example, the Advanced LIGO interferometer. Its projected circulating power (Table \ref{table:notations}) corresponds to the normalized power $\Theta/(2\pi)^3 \approx (100\,{\rm Hz})^3$ with the interferometer bandwidth $\kappa_I/2\pi\approx500\,{\rm Hz}$ \cite{CQG_32_7_074001_2015}. Tuning the Larmor frequency of spins to $\Omega_S/2\pi=3\,{\rm Hz}$, we arrive at the requirement for $\Gamma_S/2\pi\approx 600\,{\rm Hz}$.  Note that from Eq.\,\eqref{spin_noise}  we can infer that on resonance the ratio of the ground state noise contribution (last term) to the shot noise is equal to half the ratio of the QBA (second term) to the ground state noise and is given by $\Gamma_S/\gamma_S=d_0$. In \cite{Wasilewski_PRL_104_133601_2010} we achieved this ratio and hence the $d_0 \approx2$ for a single pass interaction in the atomic cells with the length of 4\,cm. Increasing the length of the cell to 10\,cm and placing the spin ensemble in an optical resonator with the finesse of $\mathcal{F}\approx70$ will provide $d_0\approx200$. The 25\,mm room temperature Caesium cells with advanced wall coating \cite{Hammerer_RMP_82_1041_2010, Balabas_OE_18_5825_2010}  have the intrinsic linewidth  $<1\,{\rm Hz}$ . With optical power broadening to $\gamma_S/2\pi=3\,{\rm Hz}$ the linewidth will be dominated by the readout and we will achieve the required value of $\Gamma_S/\gamma_S$. 

In the 10-m prototype GWD case, the best sensitivity frequency band is shifted to upper frequencies by about one order of magnitude, which relaxes requrements for $\Omega_S$, $\gamma_S$ proportionally, see Table \ref{table:notations}. In this case, the value of $\mathcal{F}\approx35$ is sufficient. 

In Figure \ref{Fig:noise}, the resulting quantum noise spectral densities calculated in SI  are shown for realistic optical losses listed in the figure caption. In the inset, we plot the susceptibility functions for a free mass and for the spin system matching the free mass over the entire frequency band of interest. It follows from these results that the proposed scheme allows to achieve the sensitivity gain about 6\,db across the entire sensitivity band of interest, which corresponds to almost an order of magnutude gain in the ``visible'' part of Universe and a proportional increase of the event rate. 

\paragraph{Conclusion.} 
 
We present a way to suppress quantum noise in gravitational wave interferometers by adding a spin system into the detection path. The proposed method allows for broadband detection sensitivity beyond the Standard Quantum Limit across the entire frequency bandwidth relevant for gravitational wave observation. In comparison to the earlier proposals for beyond the SQL GWD which use either an external filtering cavity \cite{02a1KiLeMaThVy} or utilize the GW interferometer as an effective filtering cavity \cite{Ma_NPhys_13_776_2017} our approach has an advantage of being completely compatible with existing GWDs and thus not requiring complex alterations in the GWD's core optics. QBA-evading measurement paves the road towards generation of an entangled state of the multi-kilogram GWD mirrors and atomic spins which would be of fundamental interest due to the sheer size of the objects involved.

\acknowledgments

We acknowledge motivating discussions with Nergis Mavalvala and Antonios Kontos. We acknowledge also helpful comments from Yiqui Ma. E.S.P acknowledges funding by the ERC grant INTERFACE, by ARO grant W911NF and the John Templeton Foundation. F.K. acknowledges funding by the RFBR grants 16-52-10069 and 16-52-12031.


\newpage

\section*{Supplementary information}
 
\subsubsection{Input/output relation}

\begin{table*}
  \begin{ruledtabular}
    \begin{tabular}{lll}
      Notation & Quantity & Value \\
      \hline
      $\eta_{I1}$  & quantum efficiency of the input path of the interferometer & 97.5\% \\
      $\eta_{I3}$  & quantum efficiency of the output path of the interferometer & 97.5\%\\
      $A_I$ & Light losses per bounce in the interferometer & 0.01\% \\ 
      $\kappa_{Il} = (1-\eta_{I2})\kappa_I=\dfrac{cA_I}{4L}$ & part of the interferometer half-bandwidth due to the optical losses \\
      $\kappa_{Ic} = \eta_{I2}\kappa_I$ & part of the interferometer half-bandwidth due to the coupling \\
      $\eta_{S1}$  & quantum efficiency of the input path of the spin system system & 97.5\%\\
      $\eta_{S3}$  & quantum efficiency of the output path of the atomic system & 97.5\% \\
      $A_S$ & Light losses per bounce in the spin system cavity & 0.3\% \\
      $T_S$ & Transmissivity of the coupling mirror of the spin system cavity \\[-0.5ex]
      $\eta_{S2} = \dfrac{T_S}{T_S+A_S}$ &   
    \end{tabular}
  \end{ruledtabular}
  \caption{Notation related to the optical losses}\label{table:notations2} 
\end{table*}


With optical losses taken into account, Eqs.\,(1, 4) of the main text can be shown \cite{s_12a1DaKh, s_Hammerer_RMP_82_1041_2010} to take the following form:
\begin{subequations}
  \begin{align}
    \hat{{\rm b}}_I^s =& \sqrt{\eta_{I3}}\biggl[
      \mathcal{R}_I\hat{{\rm c}}_I^s 
      + \frac{2\kappa_{Ic}\Theta\chi}{\ell^2}\,\hat{{\rm c}}_I^c
        + \mathcal{T}_I\biggl(
              \hat{{\rm n}}_{I2}^s + \frac{\Theta\chi}{\ell}\,\hat{{\rm n}}_{I2}^s
            \biggr) \nonumber \\
        &+ \sqrt{\frac{2\kappa_{Ic}\Theta}{\hbar m}}\,\frac{\chi F_s}{\ell}
      \biggr]
      + \sqrt{1-\eta_{I3}}\,\hat{{\rm n}}_{I3} \,, \label{bf_b_I} \\
    \hat{{\rm b}}_S^s =& \sqrt{\eta_{S3}}\bigl[
        \mathcal{R}_S\hat{{\rm c}}_S^s 
        + 2\eta_{S2}\theta\chi_S\hat{{\rm c}}_S^c
        + \mathcal{T}_S(\hat{{\rm n}}_{S2}^s + \theta\chi_S\hat{{\rm n}}_{S2}^c)  
          \nonumber \\
        &+ \sqrt{2\eta_{S2}\theta}\,\chi_Sf_S
      \bigr]
      + \sqrt{1-\eta_{S3}}\,\hat{{\rm n}}_{S3} \,. \label{b_S^s}
  \end{align}
\end{subequations}
Here
\begin{subequations}
  \begin{gather}
    \hat{{\bf c}}_I^{c,s} = \sqrt{\eta_{I1}}\,\hat{{\rm a}}_I^{c,s} 
      + \sqrt{1-\eta_{I1}}\,\hat{{\rm n}}_{I1}^{c,s} \,, \\
    \hat{{\bf c}}_S^{c,s} = \sqrt{\eta_{S1}}\,\hat{{\rm a}}_S^{c,s} 
      + \sqrt{1-\eta_{S1}}\,\hat{{\rm n}}_{S1}^{c,s} 
  \end{gather}
\end{subequations}
are the effective incident light quadratures for the interferometer and the spin system cavity, $\hat{{\rm n}}_{I1}^{c,s}$ and $\hat{{\bf n}}_{I3}^{c,s}$, $\hat{{\rm n}}_{S1}^{c,s}$ and $\hat{{\rm n}}_{I3}^{c,s}$ are the vacuum noise operators associated with the input and the output losses, $\hat{{\rm n}}_{S2}^{c,s}$, $\hat{{\rm n}}_{I2}^{c,s}$ are the vacuum noise operators due to the internal optical losses in the interferometer/spin system cavity, 
\begin{subequations}
  \begin{gather}
    \ell = \kappa_I - i\Omega \,, \\
    \mathcal{R}_I = \frac{2\kappa_{Ic}}{\ell} - 1 \,, \qquad 
    \mathcal{T}_I = \frac{2\sqrt{\kappa_{Ic}\kappa_{Il}}}{\ell} \,, \\
    \mathcal{R}_S = \frac{T_S - A_S}{T_S + A_S} \,, \qquad
    \mathcal{T}_S = \frac{2\sqrt{T_SA_S}}{T_S + A_S} \,.
  \end{gather}
\end{subequations}
Other notations are listed in Table \ref{table:notations2}.

It follows from \eqref{bf_b_I} that the signal force estimate without detection on the spin system is given by
\begin{equation}
  \tilde{F}_{s\,\rm I} 
  = \sqrt{\frac{\hbar m}{2\kappa_{Ic}\Theta}}\frac{\kappa_I-i\Omega}{\chi}
      \,\hat{{\rm b}}_I^s
  = F_s + \hat{F}_{\rm I} \,,
\end{equation} 
where
\begin{multline}\label{F_raw} 
  \hat{F}_{\rm I} 
    = \sqrt{\frac{\hbar m}{2\kappa_{Ic}\Theta}}\frac{\ell}{\chi}\Biggl[
      \mathcal{R}_I\hat{{\rm c}}_I^s 
      + \frac{2\kappa_{Ic}\Theta\chi}{\ell^2}\,\hat{{\rm c}}_I^c
      + \mathcal{T}_I\biggl(\hat{{\rm n}}_{I2}^s 
            + \frac{\Theta\chi}{\ell}\hat{{\rm n}}_{I2}^c
          \biggr) \\  
      + \sqrt{\frac{1-\eta_{I3}}{\eta_{I3}}}\,\hat{{\rm n}}_{I3} 
    \Biggr]  
\end{multline}
is the quantum noise in the interferometer channel alone.

\subsubsection{Spectral densities}

Spectral densities of the input noise components [see Eq.\,(9) of the main text] are defined by the degree of two-mode squeezing
\begin{subequations}\label{S_a} 
  \begin{equation}
    S[\hat{{\rm a}}_I^c] = S[\hat{{\rm a}}_I^s] 
    = S[\hat{{\rm a}}_S^c] = S[\hat{{\rm a}}_S^s] = \frac{\cosh2r}{2} \,.
  \end{equation} 
  The only non-zero cross-correlation spectral densities are
  \begin{equation}
    S[\hat{{\rm a}}_I^c\hat{{\rm a}}_S^c] = \frac{\sinh2r}{2}\,, \qquad 
    S[\hat{{\rm a}}_I^s\hat{{\rm a}}_S^s] = -\frac{\sinh2r}{2} \,.
  \end{equation} 
\end{subequations}
Therefore, spectral densities of $\hat{{\rm c}}_{I,S}^{c,s}$ and their (non-zero) cross-correlation spectral densities are equal to, respectively
\begin{subequations}\label{S_c} 
  \begin{gather}
    S[\hat{{\rm c}}_I^c] = S[\hat{{\rm c}}_I^s] = \frac{\rho_I+1}{2} \,, \\
    S[\hat{{\rm c}}_S^c] = S[\hat{{\rm c}}_S^s] = \frac{\rho_S+1}{2} \,, \\
    S[\hat{{\rm c}}_I^c\hat{{\rm c}}_S^c] = \frac{\rho_{IS}}{2} \,, \qquad 
      S[\hat{{\rm c}}_I^s\hat{{\rm c}}_S^s] = -\frac{\rho_{IS}}{2} \,, 
  \end{gather}
\end{subequations}
where
\begin{subequations}
  \begin{gather}
    \rho_I = 2\eta_{I1}\sinh^2r \,, \\ 
    \rho_S = 2\eta_{S1}\sinh^2r \,, \\ 
    \rho_{IS} = \sqrt{\eta_{I1}\eta_{S1}}\sinh2r \,.
  \end{gather}
\end{subequations}
The force spectral densities for the interferometer and the spin system (\ref{F_raw}, \ref{b_S^s}) and their cross-correlation spectral density are:
\begin{subequations}
  \begin{gather}
    S_{\rm I} = \frac{\hbar m|\ell|^2}{4\kappa_{Ic}\Theta|\chi|^2}\,\sigma_{\rm I}\,,
      \\
    S_{\rm Spin} = \frac{\eta_{S3}}{2}\,\sigma_{\rm Spin} \,, \\
    S_{\rm I\,Spin} = \frac{1}{2}\sqrt{\frac{\hbar m\eta_{S3}}{2\kappa_{Ic}\Theta}}\,
      \frac{\ell}{\chi}\,\sigma_{\rm I\,Spin} \,.
  \end{gather}
\end{subequations}
Here 
\begin{subequations}\label{sigmas} 
  \begin{gather}
    \sigma_{\rm I} = |\mathcal{R}_I|^2\rho_I + \frac{1}{\eta_{I3}} 
      + \frac{4\eta_{I2}\kappa_I^2\Theta^2|\chi|^2}{|\ell|^4}(\eta_{I2}\rho_I + 1), \\
    \sigma_{\rm Spin} = \mathcal{R}_S^2\rho_S + \frac{1}{\eta_{S3}} 
      + 4\eta_{S2}\theta^2|\chi_S|^2(\eta_{S2}\rho_S+1) \nonumber \\
        + 4\eta_{S2}\theta|\Im\chi_S| \,, \\
    \sigma_{\rm I\,Spin} = \left(
        -\mathcal{R}_I\mathcal{R}_S 
        + \frac{4\kappa_{Ic}\eta_{S2}\Theta\theta\chi\chi_S^*}{\ell^2}
      \right)\rho_{IS} \,.  
  \end{gather}
\end{subequations}
The force estimate for the hybrid interferometer-spin system is equal to
\begin{equation}
  \tilde{F}_s = \tilde{F}_{s\,\rm I} + \alpha\hat{{\rm b}}_S^s \,,
\end{equation} 
where
\begin{equation}
  \alpha = -\frac{S_{\rm I\,Spin}}{S_{\rm Spin}} \,.
\end{equation} 
The resulting effective position noise spectral density is equal to
\begin{equation}
  S = |\chi|^2\left(S_{\rm I} - \frac{|S_{\rm I\,Spin}|^2}{S_{\rm Spin}}\right)
  = \frac{\hbar m|\ell|^2}{4\kappa_{Ic}\Theta}\,
      \biggl(\sigma_{\rm I} - \frac{|\sigma_{\rm I\,Spin}|^2}{\sigma_{\rm Spin}}\biggr)
    \,.
\end{equation} 

\subsubsection{Balancing of the back actions}

Consider the leading in $e^r$ terms in Eqs.\,\eqref{sigmas}. In this approximation,
\begin{subequations}
  \begin{gather}
    \rho_I \approx \frac{\eta_{I1}e^{2r}}{2} \,, \\ 
    \rho_S \approx \frac{\eta_{S1}e^{2r}}{2} \,, \\ 
    \rho_{IS} \approx \frac{\sqrt{\eta_{I1}\eta_{S1}}\,e^{2r}}{2} \,,
  \end{gather}
\end{subequations}
\begin{subequations}
  \begin{gather}
    \sigma_{\rm I} \approx \frac{\eta_{I1}e^{2r}}{2}
      \biggl(|\mathcal{R}_I|^2 + \frac{4\eta_{I2}^2\kappa_I^2\Theta^2|\chi|^2}
        {|\ell|^4}\biggr),\\
    \sigma_{\rm Spin} \approx \frac{\eta_{S1}e^{2r}}{2}\biggl(
        \mathcal{R}_S^2 + 4\eta_{S2}^2\theta^2|\chi_S|^2
      \biggr) , \\
    \sigma_{\rm I\,Spin} \approx \frac{\sqrt{\eta_{I1}\eta_{S1}}\,e^{2r}}{2}\biggl(
        -\mathcal{R}_I\mathcal{R}_S 
        + \frac{4\eta_{I2}\eta_{S2}\kappa_I\Theta\theta\chi\chi_S^*}{\ell^2}
      \biggr) .
  \end{gather}
\end{subequations}
and
\begin{equation}
  \sigma_{\rm I} - \frac{|\sigma_{\rm I\,Spin}|^2}{\sigma_{\rm Spin}}
  \approx \frac{\eta_{I1}e^{2r}}{2}\frac{\left|
      \dfrac{\mathcal{R}_S\eta_{I2}\kappa_I\Theta\chi}{\ell^2} 
      + \mathcal{R}_I\eta_{S2}\theta\chi_S
    \right|^2}
    {\mathcal{R}_S^2 + 4\eta_{S2}^2\theta^2|\chi_S|^2} \,.  
\end{equation} 
In order to compensate the noise, the numerator of this equation has to be equal to zero.  The simplified frequency-independent form of this condition is the following:
\begin{equation}\label{theta_cond} 
  \frac{\eta_{S2}}{2\eta_{S2}-1}\,\theta
  = \frac{\eta_{I2}}{2\eta_{I2}-1}\,\frac{\Theta}{\kappa_I} \,.
\end{equation} 
Taking into account that 
\begin{equation}
  \theta = \frac{4}{T_S+A_S}\,\theta_{\rm SP} = \frac{2\mathcal{F}}{\pi}\,\theta_{\rm SP}
  \,,
\end{equation} 
where $\theta_{\rm SP}$ is the single-pass value of the coupling factor $\theta$, Eq.\,\eqref{theta_cond} can be presented as follows:
\begin{equation}
  T_S^2 -2bT_S - A_S^2 = 0 \,,
\end{equation} 
where
\begin{equation}
  b = \frac{2\eta_{I2}-1}{\eta_{I2}}\,\frac{2\kappa_I\theta_{\rm SP}}{\Theta} \,.
\end{equation} 
Therefore, for given $\theta_{\rm SP}$ and $A_S$, the back actions balancing is provided by  
\begin{equation}
  T_S = b + \sqrt{b^2 + A_S^2} \,.
\end{equation}

\end{document}